\documentclass[aps,twocolumn,superscriptaddress,showpacs,footnote]{revtex4}

\usepackage{graphicx,pstricks}
\def\lsim{\stackrel{\scriptstyle <}{\phantom{}_{\sim}}}
\def\gsim{\stackrel{\scriptstyle >}{\phantom{}_{\sim}}}
\newcommand{\be}{\begin{equation}}
\newcommand{\ee}{\end{equation}}
\newcommand{\ba}{\begin{eqnarray}}
\newcommand{\ea}{\end{eqnarray}}

\begin{document}

\title{Nuclear medium cooling scenario
in the light of new Cas A cooling data and \\
the $2~M_\odot$ pulsar mass measurements}

\author{D. Blaschke}
  \affiliation{Institute for Theoretical Physics,
    University of Wroc{\l}aw, 50-204 Wroc{\l}aw, Poland}
  \affiliation{Bogoliubov Laboratory for Theoretical Physics, Joint
    Institute for Nuclear Research, 141980 Dubna, Russia}

\author{H. Grigorian}
 \affiliation{Department of Theoretical
    Physics, Yerevan State University, %Alex Manoogyan 1,
    375025 Yerevan, Armenia}
 \affiliation{Laboratory for Information Technologies, Joint
    Institute for Nuclear Research, 141980 Dubna, Russia}

\author{D. N. Voskresensky}
 \affiliation{ National Research Nuclear
    University (MEPhI), 115409 Moscow, Russia}

\date{\today}

\begin{abstract}
Recently, Elshamounty et al.  performed a reanalysis of the
surface temperature of the neutron star in the supernova remnant
Cassiopeia A on the basis of Chandra data measured during last
decade, and added a new data point. We show that all reliably
known temperature data of neutron stars including those belonging
to Cassiopea A can be comfortably explained in our "nuclear medium
cooling" scenario of neutron stars. The cooling rates account  for
medium-modified one-pion exchange in dense matter, polarization
effects in the pair-breaking-formation processes operating on
superfluid  neutrons and protons paired in the $1S_0$ state, and
other relevant processes. The emissivity of the
pair-breaking-formation process in the $3P_2$ state is a tiny
quantity within our scenario. Crucial for a successful description
of the Cassiopeia A cooling proves to be the thermal conductivity
from both, the electrons and nucleons, being reduced by medium
effects. Moreover, we exploit an EoS which stiffens at high
densities due to an excluded volume effect and is capable of
describing a maximum mass of $2.1~M_\odot$, thus including the
recent measurements of PSR J1614-2230 and PSR J0348+0432.

\end{abstract}


% 97.60.Jd   Neutron stars
% 95.30.Cq   Elementary particle processes
% 26.60.-c   Nuclear matter aspects of neutron stars

\maketitle

\section{Introduction}
%{\it Introduction.--}
The isolated neutron star in Cassiopeia~A (Cas~A) was discovered in
1999 by the {\em Chandra} satellite \cite{Tananbaum:1999kx}.  Its
association with the historical supernova SN~1680
\cite{Ashworth:1980vn} gives Cas~A an age of 333~years, in agreement
with the nebula's kinematic age \cite{Fesen:2006ys}.
The thermal soft X-ray spectrum of Cas~A can be fitted with a
non-magnetized carbon atmosphere model, a surface temperature of
$2\times 10^6$~K, and an emitting radius of 8 to 17~km \cite{Ho:2009fk}.
Analyzing the data from 2000 to 2009, Heinke \& Ho \cite{Heinke:2010xy}
reported a rapid decrease of Cas~A's surface temperature over a 10-year
period, from $2.12 \times 10^6$ to $2.04 \times 10^6$~K.
Such a rapid drop in temperature conflicts with standard cooling scenarios
based on the efficient modified Urca (MU) process
~\cite{Yakovlev:2000jp,Page:2006ly}.
Interpretations of Cas~A's temperature data based on hadronic matter
cooling scenarios were provided  by Page {\it et al.} \cite{Page:2010aw},
Yakovlev {\it et al.} \cite{Shternin:2010qi,Yakovlev:2010ed},
Blaschke {\it et al.} \cite{Blaschke:2011gc}.
The new analysis of the {\it Chandra} data performed by
Elshamounty {\it et al.} \cite{Elshamouty:2013nfa} including a new measured
data point allowed to precisely extract the decade temperature decline.
The drop in the temperature lies in the range $2\dots 5.5\%$, and the most
recent results from ACIS-S detector yield a $3\dots 4\%$ decline.

The interpretation of Cas A data by Page {\it et al.}
\cite{Page:2010aw} is based on the ``minimal cooling'' paradigm
\cite{Page:2004fy}, where a minimal number of cooling processes is
taken into account. These are photon emission, the  MU process,
nucleon-nucleon ($NN$) bremsstrahlung (NB) and the neutron ($n$)
and proton ($p$) pair breaking-formation processes (nPBF and
pPBF). To calculate the $NN$ interaction entering the emissivities
of the MU and NB processes the minimal cooling scenario employs
the free one-pion exchange (FOPE) model \cite{Friman:1978zq}. As
shown in \cite{Page:2010aw}, the Cas~A data can be  reproduced by
assuming a large value for the proton pairing gap throughout the
entire stellar core. The latter assumption facilitates additional
suppression of the emissivity of the MU  and the pPBF processes.
Under this assumption the nPBF reaction in the core, where
neutrons are paired in $3P_2$ state, proves to be the most
efficient one. The authors  describe Cas A data by fixing the
critical temperature for the neutron $3P_2$ pairing gap at around
$0.5\times 10^9$~K. The result is mildly sensitive to the neutron
star mass. Surface temperature--age data of other neutron stars,
which do not lie on the cooling curve of Cas~A, are explained
within the minimal cooling scenario mainly by assuming variations
in the light element mass of the envelopes of these stars. However
both, younger neutron stars like the one in CTA1, and very old
hotter stars require more than minimal cooling \cite{Page:2012se}.

The works of Yakovlev {\it et al.}
\cite{Shternin:2010qi,Yakovlev:2010ed} include all emission
processes which are part of the minimal cooling paradigm including
the FOPE  model for MU reaction rate. They assume that the proton
gap is so large that charged current processes are strongly
suppressed in the entire stellar core. The value and the density
dependence of the $3P_2$ neutron gap are fitted to the Cas~A data,
leading to a critical temperature of $0.7\dots 0.9\times 10^9$~K
for the neutron pairing gap and a neutron star mass
$M=1.65~M_{\odot}$. Both groups therefore came to the striking
conclusion that the temperature data of Cas~A allow one to extract
the value of the $3P_2$ neutron pairing gap. Continuing this
approach Elshamouty {\it et al.} \cite{Elshamouty:2013nfa} arrive
at the same conclusion.

The work of Blaschke  {\it et al.} \cite{Blaschke:2011gc} presents
the ``nuclear medium cooling scenario''  as a model for the successful
description of all the known temperature data including those of Cas~A.
This scenario includes efficient medium modified Urca (MMU)  and
medium nucleon bremstrahlung (MnB, MpB) processes, as motivated by a softening
of the virtual pion mode in dense matter \cite{Migdal:1978az,Migdal:1990vm},
a very low (almost zero) value of the $3P_2$ neutron gap, as motivated by
the result of Schwenk and Friman \cite{Schwenk:2003bc}, and a small
thermal conductivity of neutron star matter caused by  in-medium
effects, as motivated by calculations of the lepton thermal
conductivity of Shternin and Yakovlev \cite{Shternin:2007ee} and evaluations
of the effect of pion softening on the nucleon thermal conductivity.
More specifically, in \cite{Blaschke:2011gc} we just
used values of the thermal conductivity calculated by Baiko  {\it et al.}
\cite{Baiko:2001cj} suppressed by a parameter $\zeta_{\kappa}$.
The best fit of Cas A data was performed for $\zeta_{\kappa}= 0.265$.
A strong suppression of the thermal conductivity is justified by results of
Shternin and Yakovlev \cite{Shternin:2007ee}, who included the in-medium effect
of Landau damping of electromagnetic interactions owing to the
exchange of transverse plasmons in the partial  electron (and
muon) contribution to the thermal conductivity.
Earlier, this effect has been studied by Heiselberg and Pethick for a
degenerate quark plasma \cite{Heiselberg:1993cr} and  by Jaikumar
{\it et al.} \cite{Jaikumar:2005gm} for neutrino bremsstrahlung radiation
via electron-electron collisions in neutron star crusts and cores.
Now, we incorporate the in-medium modifications of the electron-electron
interaction into our scenario, precisely as it has been done in
\cite{Shternin:2007ee}.
Moreover, the partial $NN$ thermal conductivity should be suppressed within
our scenario owing to the increase of the squared
$NN$ interaction matrix element with density caused by the medium modification
of the FOPE.
Thereby, we additionally suppress the $NN$ thermal conductivity term
calculated in \cite{Baiko:2001cj} by taking into account the softening of the
one-pion exchange for this quantity as well as for all processes considered in
our scenario.

As the nuclear matter equation of state (EoS), in \cite{Blaschke:2011gc}
we used  the Heiselberg-Hjorth-Jensen (HHJ) EoS \cite{Heiselberg:1999fe}
(with a fitting parameter $\delta =0.2$) that fits the microscopic
Akmal-Pandharipande-Ravenhall (APR) $A18+\delta v+UIX^*$ EoS
\cite{Akmal:1998qf} for symmetric nuclear matter up to $4n_0$,
 where $n_0=0.16$~fm$^{-3}$ is the nuclear saturation density.
This yields an acceptable (although not perfect) fit of the APR
EoS of neutron star matter for those densities. The maximum
neutron star mass calculated with the HHJ($\delta =0.2$) EoS,
$M_{\rm max}=1.94~M_{\odot}$,  proves to be smaller than the one
calculated with the original APR EoS, $M_{\rm max}\simeq
2.2~M_{\odot}$. However, the latter EoS becomes acausal for
$n>0.86$~fm$^{-3}$, whereas all HHJ($\delta \geq 0.13$) EoS
%{\red
respect
%}
causality at all densities. Recent measurements of two
massive neutron stars, with $M_{1614}=1.97\pm 0.04~M_{\odot}$ for
PSR J1614-2230 \cite{Demorest:2010bx} and $M_{0348}=2.01\pm
0.04~M_{\odot}$ for PSR J0348-0432 \cite{Antoniadis:2013pzd},
motivate us to use a stiffer EoS than that of HHJ($\delta =0.2$)
at large densities. In the present work we modify the EoS in order
to fulfill the new observational constraints on masses of neutron
stars. To that end we incorporate excluded volume corrections in
the HHJ($\delta =0.2$) EoS such that it would remain unchanged for
$n\lsim 4 n_0$ but would become stiffer for higher densities.

Thus our  aim in the given work is to demonstrate the efficiency of
our nuclear medium cooling scenario in explaining the cooling data
introducing the lepton contribution to the thermal conductivity
following \cite{Shternin:2007ee} and  extending the HHJ($\delta=0.2$)
EoS to describe the new data on massive stars.

\section{Nuclear medium cooling scenario}
The nuclear medium cooling scenario worked out in
Refs.~\cite{Voskresensky:1986af,Voskresensky:1987hm,
Migdal:1990vm,Voskresensky:2001fd,Kolomeitsev:2010pm} has been
successfully applied to the description of the body of known
surface temperature--age data of neutron stars
\cite{Schaab:1996gd,Blaschke:2004vq,Grigorian:2005fn,Grigorian:2005fd,Blaschke:2011gc}.
It exploits a strong dependence of the main cooling mechanisms on
the density and thus on the neutron star mass.

\subsection{Free versus medium-modified one-pion-exchange in dense matter}
We exploit the Fermi liquid approach, where the short-range
interaction is treated with the help of phenomenological
Landau-Migdal parameters, whereas long-range collective modes are
explicitly  presented. The most important effect comes from the
mode  with the pion quantum numbers treated explicitly, as it is a
soft mode ($m_{\pi}\ll m_N$, with $m_{\pi}$ $(m_N)$ being the pion
(nucleon) mass). The key effect is the softening of the pion mode
with increasing density \cite{Migdal:1978az,Migdal:1990vm}. Only
with the inclusion of this softening effect the phase transition
to a pion condensation state in dense nucleon matter may appear.
Thus it is quite inconsistent to use FOPE model for description of
$NN$-interaction and simultaneously include processes going on
pion condensation.

The insufficiency of the FOPE model for the description of the
$NN$-interaction is a known issue. Actually, using the FOPE for
the $NN$ interaction amplitude, and simultaneously considering
pion propagation as free,  violates unitarity. Indeed, calculating
the MU emissivity perturbatively one may use both the Born $NN$
interaction amplitude given by the FOPE and the optical theorem,
considering the imaginary part of the pion self-energy
\cite{Voskresensky:1986af,Voskresensky:1987hm,Voskresensky:2001fd}.
In the latter case, at low densities one needs to expand the exact
pion Green's function $D_{\pi}(\omega,k)=[\omega^2 -m_{\pi}^2 -k^2
-\Pi(\omega,k,n)]^{-1}$ to second order using for the polarization
function $\Pi(\omega,k,n)$ the perturbative one-loop diagram,
$\Pi_0(\omega,k,n)$. For $k=k_0$, which is the pion momentum at
the minimum of the effective pion gap defined as
$\omega^{*\,2}=-D_{\pi}^{-1}(\omega =0,k =k_0)$, the polarization
function $\Pi_0 (\omega,k=k_0\simeq p_{{\rm F},n},n)$ yields  a
strong  $P$-wave attraction. Here $p_{{\rm F},n}$ is the neutron
Fermi momentum. This attraction proves to be so strong that it
would trigger a pion condensation instability already at low
baryon densities of $n \sim 0.3\, n_0$, which is in disagreement
with experimental data on atomic nuclei at the nuclear saturation
density $n_0$.
% = 0.16$~fm$^{-3}$.
Note that the perturbative calculation contains no one free
parameter. The paradox is resolved by observing that together with
pion softening (i.e., a decrease of the effective pion  gap
$\omega^{*}(n)$ with increasing density for $n> n_{cr}^{(1)}$,
$\omega^{*\,2}(n_{cr}^{(1)})= m_{\pi}^2$) one needs to include a
short-range repulsion arising from the dressed $\pi NN$ vertices,
$\Gamma (n)\simeq [1+C (n/n_0)^{1/3}]^{-1}$ with $C\simeq 1.6$.
This evaluation exploits an estimated value of the Fermi-liquid
spin-spin Landau-Migdal parameter $g$ and the Lindhard function
taken in the limit of low transferred energy $\omega \ll p_{{\rm
F},n}$. A consistent description of the $NN$ interaction in matter
should thus use a medium modified one-pion exchange (MOPE)
interaction characterized by the fully dressed pion Green
function, dressed vertices $\Gamma(n)$, and a residual $NN$
interaction.  We stress that dressing the pion mode is similar to
the ordinary dressing of the photon mode in a plasma. Computing
similar diagrams results in a dispersion of the dielectric
constant $\varepsilon (\omega, q)\neq 1$, i.e. it might
essentially deviate from unity. Moreover, only by including
dressed vertices one is able to describe zero sound modes in Fermi
liquids. Microscopic calculations of the residual interaction are
very cumbersome.
However, according to  evaluations in \cite{%Migdal:1978az,
Voskresensky:1986af,Migdal:1990vm}, the main contribution for
$n>n_0$ is given by MOPE, whereas the relative contribution of the
residual interaction diminishes  with increasing density owing to
polarization effects. Thus, in our simplified treatment the main
dependence on the short-range interaction  enters MOPE  via the
phenomenological vertex suppression factor %$\Gamma (C(g))$.
$\Gamma (n)$.

\begin{figure}[!tbh]
\includegraphics[width=0.5\textwidth]{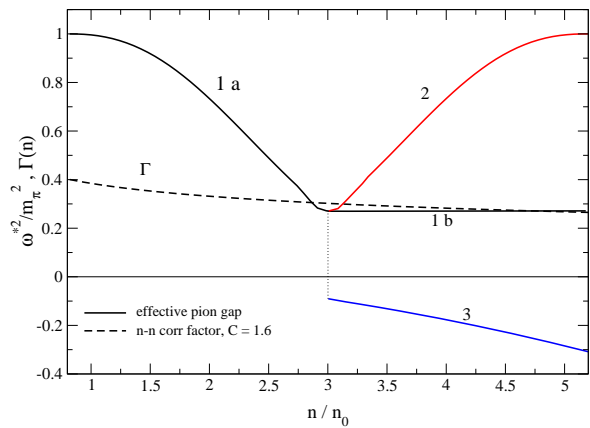}
\caption{(Color online) 
Square of the effective pion gap $\omega^*$ with pion
condensation (branches 1a+2 and 3) and without (1a+ 1b).
$\Gamma(n)$ is the  nucleon - nucleon correlation factor.
%[omegatil.ps]
\label{omegatil} }
\end{figure}

The density dependence of the effective pion gap $\omega^*$ that
we use for $n>n_{cr}^{(1)}$, taken to be $0.8~n_0$, is
demonstrated in Fig.~\ref{omegatil} (Fig.~1 of
\cite{Blaschke:2004vq}). The curve 1a in Fig.~\ref{omegatil} shows
behavior of the pion gap for $n<n_{cr}^{\pi}$, where
$n_{cr}^{\pi}$, taken to be $3 n_0$,  is the critical density for
the pion condensation. For simplicity we do not distinguish
between different possibilities of $\pi^0$, $\pi^{\pm}$
condensations, see \cite{Migdal:1990vm} for a more general
description. Note that variational calculations of Akmal {\it et
al.} \cite{Akmal:1998qf} produce still smaller critical densities
for the charged and neutral pion condensations. However, in order
to be conservative we assume a larger value of $n_{cr}^{\pi}$.
Following the model used here and in
\cite{Blaschke:2004vq,Grigorian:2005fn} within the
HHJ($\delta=0.2)$  EoS, the pion condensation arises for neutron
star masses $M\geq 1.32~M_{\odot}$ (corresponding to the choice
$n\geq n_{cr}^{\pi}= 3\, n_0$). The curve 1b demonstrates the
possibility of a saturation of pion softening and absence of the
pion condensation for $n>n_{cr}^{\pi}$ (this possibility could be
realized, e.g., if the Landau-Migdal parameters increased with the
density). Thus the curves 1a$+$1b  determine behavior of the Green
function for the pion excitations in absence of condensation.
Curves 2, 3 demonstrate  possibility of the pion condensation for
$n>n_{cr}^{\pi}$. The continuation of the branch 1a for
$n>n_{cr}^{\pi}$, called  branch 2, shows the reconstruction of
the pion dispersion relation in the presence of the condensate
state.  In the presence of the pion condensate (for
$n>n_{cr}^{\pi}$) the value $\omega^*$ from curve 2 enters the
emissivities of all processes with pion excitations in initial,
intermediate and final reaction states. In agreement with the
general trend known in condensed matter physics, fluctuations
dominate in the vicinity of the critical point of the phase
transition (where $\omega^*$ has its smallest values) and die out
far away from it. In strongly interacting systems, like $^4$He,
fluctuations prove to be important at all temperatures. The jump
from branch 1a to branch 3 at $n=n_{cr}^{\pi}$ is due to the first
order phase transition to the $\pi$ condensation, see
\cite{Migdal:1990vm}. The $|\omega^*|$ value on branch 3 is
proportional to the amplitude of the pion condensate mean field.
To avoid misunderstanding we stress that, although to construct
the curves $\omega^* (n)$ we used available experimental
information and well established general principles
\cite{Migdal:1990vm}, the quantitative density dependence of
$\omega^* (n)$ remains essentially model dependent due to a
lacking knowledge of the $NN$ interaction in neutron star matter
at large densities. Thus we hope that our successful description
of the neutron star cooling may be helpful to correctly choose the
parameterization of the interaction.

%%%%%%%%%%%%%%%%%  Figure 1 %%%%%%%%%%%%%%%
%%%%%%%%%%%%%%%%%%%%%%%%%%%%%%%%%%%%%%%%%%

The direct Urca (DU) process, $n\to pe\bar{\nu}$, is too efficient
for a description of the full set of neutron star cooling data.
Moreover, the DU process occurs only for high proton fractions,
$x_p = n_p/n > 0.11\dots 0.14$.
In the calculation of the neutrino emissivity of two-nucleon processes,
e.g., $nn\to npe\bar{\nu}$, not only radiation from the nucleon legs but
also from intermediate reaction states is allowed.
For $n\gsim n_0$, the latter processes prove to be more efficient than the
ordinary MU process from the legs.
With such an interaction the ratio of the emissivity of the
medium modified Urca (MMU) to the MU process, see
\cite{Blaschke:2004vq,Grigorian:2005fn,Blaschke:2011gc},
\be \frac{\epsilon_{\nu}[\rm MMU]}{\epsilon_{\nu}[\rm MU]}\sim
3\left(\frac{n}{n_0}\right)^{10/3}
\frac{[\Gamma(n)/\Gamma(n_0)]^6}{[\omega^{*}(n)/m_{\pi}]^8} \, ,  \ee
strongly increases with density for $n\gsim n_0$. For
$n<n_{cr}^{(1)}$ we use $\epsilon_{\nu}[\rm MU]$ as in the minimal
cooling scenario. Although an increase of the ratio of
emissivities of the medium modified nucleon (neutron)
bremsstrahlung process (MnB) to the unmodified bremsstrahlung (nB)
is less pronounced, the MnB process, being not affected by the
proton superconductivity, may yield a relatively large
contribution in the region of a strong proton pairing. Note that
being computed with values $\omega^{*}$ and $\Gamma$, which we
use, the ratio of the MOPE NN cross section to that of the FOPE
\cite{Voskresensky:2001fd} proves to be $\sigma
[\rm{MOPE}]/\sigma[\rm{FOPE}] \sim 1/3\dots 1/2$ for $n=n_0$ but
it increases with increasing density. The subsequent increase of
the cross section with density is due to the dominance of the
softening of the pion mode owing to $\pi NN$ and $\pi
N\Delta^{*}(1236)$ $P$-wave attraction compared to the suppression
of vertices owing to repulsive $NN$ correlations
\cite{Migdal:1990vm,Voskresensky:2001fd}. Thus the known
suppression of the in-medium $NN$ cross section at $n\lsim n_0$
compared to that given by the FOPE
\cite{Blaschke:1995va,Hanhart:2000ae} does not conflict with a
strong enhancement of the MMU emissivity with increasing density.
 Estimated strong density dependence of the
in-medium neutrino-processes motivated authors of \cite{
Voskresensky:1986af} to suggest that difference in surface
temperatures of neutron stars  is explained by  different masses
of the objects (that time only upper limits on surface
temperatures were put).
%{\red
At the end, we should stress that in order to explain the
cooling of both slowly  and rapidly cooling stars
one requires neutrino emissivities that differ by a factor
$>10^3$.
Therefore, an uncertainty of the order of one
in the emissivity of the processes does not affect the general
cooling picture.
%}

\subsection{Gaps and pair-breaking-formation}
\begin{figure}[!tbh]
\includegraphics[width=0.45\textwidth]{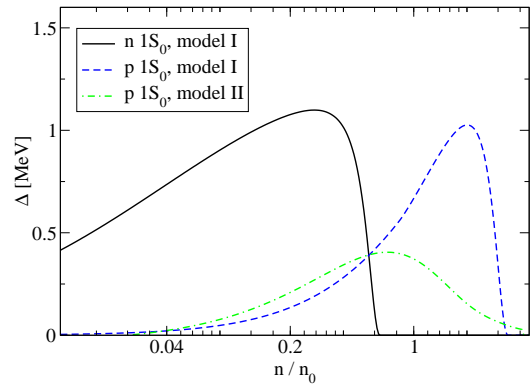}
\caption{(Color online) Neutron and proton $1S_0$ pairing gaps.
The $1S_0$ neutron gap is shown as the bold solid line. The $1S_0$
proton gap is given for model I (dashed line) and for model II
(dash-dotted line). The tiny value of the $3P_2$ gap is not shown.
%%% according to model I
%%\cite{YGKLP03}
%%%(thick solid, dashed and dotted lines) and according to model II
%%\cite{TT04}
%%%{SVSWW97}
%%%(thin lines), see text. The $1S_0$ neutron gap is the same in both
%%%models, taken from  \cite{AWP}.
\label{Gaps_2}}
\end{figure}

In spite of  many calculations performed so far, the values of
nucleon gaps in dense neutron star  matter remain poorly known.
This is a consequence of the exponential dependence of the gaps on
the potential of the in-medium $NN$ interaction. The latter
potential is not sufficiently well known. Gaps that we have
adopted in the framework of the nuclear medium cooling scenario
are presented in Fig.~\ref{Gaps_2} (cf. Fig.~5 of
\cite{Blaschke:2004vq}).

%%%%%%%%%%%%%%%%%%%% Figure 2 %%%%%%%%%%%%%%%%%%
%%%%%%%%%%%%%%%%%%%%%%%%%%%%%%%%%%%%%%%%%%%%%%%%

The $\Delta_{nn} (^1S_0)$ neutron gap is taken from \cite{APW}. As
follows from the analysis of \cite{Grigorian:2005fn} the cooling
proved to be not much sensitive to the value and density
dependence of the $1S_0$ neutron gap
%{\red
mainly since the pairing is restricted to the region of rather low
densities.
%}
Two different models \cite{Blaschke:2004vq,Grigorian:2005fn}, labeled
I and II, are used for the proton gap $\Delta_{pp} (^1S_0)$, which
is spread up to larger densities. Model I is a fit from
\cite{Yakovlev:2003qy} and model II is a calculation from
\cite{Takatsuka:2004zq}.
Refs.~\cite{Grigorian:2005fn,Blaschke:2011gc} demonstrate a strong
sensitivity of the cooling to the values and the density
dependence of the proton gap.
%{\red
For $n\lsim 0.8~n_0$
%}
neutrons are paired in the $1S_0$ state and for larger $n$, in the
$3P_2$ state.
For densities up to $3\dots 4~ n_0$ protons are paired in the $1S_0$ state.

An important in-medium effect which we incorporated in our nuclear
medium cooling scenario is the very strong suppression of the
neutron $3P_2$ pairing gap $\Delta_{nn} (3P_2)$, as  motivated by
detailed calculations of Ref.~\cite{Schwenk:2003bc}. According to
this analysis, with due account of polarization effects the gap
turns out to be $\Delta_{nn} (3P_2)\lsim $~few keV, i.e., it is
dramatically suppressed compared to BCS based calculations
\cite{Takatsuka:2004zq}. Ref.~\cite{Grigorian:2005fn} exploiting
various simulations concluded that within our cooling scenario the
$3P_2$ gap should be suppressed compared to values $(0.2\dots
0.5)\cdot 10^9$K  used by \cite{Yakovlev:2003qy,Takatsuka:2004zq}.
Thus we adopt a tiny $\Delta_{nn} (3P_2)$ gap following
\cite{Schwenk:2003bc}. Actually, the  value of the gap that yields
the best fit of Cas A data within our scenario is so tiny
\cite{Grigorian:2005fn} that it does not affect the cooling
evolution. Therefore, we do not show $\Delta_{nn} (3P_2)$ in
Fig.~\ref{Gaps_2}.

Furthermore, as it is commonly accepted, the neutron and proton
superfluidities
%with density dependent pairing gaps
cause an exponential suppression of neutrino emissivities of the
nucleon processes and of the nucleon specific heat and open up the
new class of nPBF and pPBF processes. The nPBF neutrino process
was introduced in \cite{Flowers:1976vn,Voskresensky:1987hm}, the
pPBF one, in \cite{Voskresensky:1987hm}. Their important effect on
the cooling was first incorporated in a cooling code in
Ref.~\cite{Schaab:1996gd}. Afterwards, these processes were used
in all relevant cooling codes. The important role of polarization
effects in pPBF and nPBF processes was first noted in
\cite{Voskresensky:1987hm}. Detailed analyses of the vector
current conservation in PBF reactions
\cite{Leinson:2006gf,Kolomeitsev:2008mc} have shown that by taking
the in-medium dressing of vertices into account, as it is required
by the Ward-Takahashi identities, the emissivity of processes on
the vector current proves to be dramatically suppressed $\propto
v_{\rm F}^4$, where $v_{\rm F}$ is the Fermi velocity.
Consequently, the main contribution to the PBF emissivity comes
from processes on the axial current, suppressed only as $\propto
v_{\rm F}^2$, \cite{Kolomeitsev:2008mc}. For the ratio of the
proton to the neutron PBF emissivities
%{\red
we may estimate
%}
\cite{Kolomeitsev:2008mc}
 \be R[p/n] \sim x_p^{4/3}
 (\Delta_p/\Delta_n)^{13/2}e^{2(\Delta_n -\Delta_p)/T}
 \ee
in the low temperature limit where $T\ll T_{cp}, T_{cn}$. Here
$\Delta_n$ and $\Delta_p$ are the neutron and proton $1S_0$ gaps
while $T_{cp}$ and $T_{cn}$ are corresponding critical
temperatures.
%{\red
The emissivities of the PBF processes are computed following expressions
given in Ref.~\cite{Kolomeitsev:2008mc}.
The emissivities of the two-nucleon processes are suppressed in
the presence of the pairing by the $R$-factors presented, e.g., in
Ref.~\cite{Yakovlev:2000jp}.
%}

%{\red
Concluding this subsection we stress that there are many
calculations of the $1S_0$ neutron and proton gaps and there are
some evaluations of $3P_2$ neutron gaps, for references see, e.g.,
\cite{Shen:2002pm,Zuo:2004mc,Zhou:2004fz,Baldo:2007jx,Zuo:2006gn,Zuo:2008zza,Dong:2013sqa}.
We used the same two choices of the gaps as in our previous works
\cite{Blaschke:2004vq,Grigorian:2005fn} just to demonstrate the
efficiency of the predictions of our old model. If we used results
of other calculations of $1S_0$ neutron and proton gaps we could
explain the cooling data tuning some not well known parameters,
such as the quantities $\Gamma$ and $\widetilde{\omega}$ entering
the MMU emissivity. The results are sensitive to the value and the
density dependence  of the $3P_2$ neutron gap mainly because it
spreads to large densities. Ref.~\cite{Khodel} discussed the
interesting possibility of a large $3P_2$ neutron gap (exceeding
$\sim 1$ MeV). With such a gap the region with $n>n_c ({ 3P_2})$
would be excluded from the cooling due to a strong (exponential)
suppression of the emissivities of the processes for $T\lsim 10^9$
K of our interest. Usually, one considers $3P_2$ pairing to occur
for $n\gsim n_0$. Such densities are already reached in the center
of a neutron star with mass $M\sim 0.3~M_{\odot}$, see
Fig.~\ref{HHJ-NLW-DU0nsets} below, whereas we have no observations
of neutron star masses lower than $\sim 1.1~M_{\odot}$. Thereby,
and as we have shown in \cite{Grigorian:2005fn}, within our
scenario we could not succeed to get an appropriate fit of the
data, if $3P_2$ neutron gaps were large in a broad  density
interval.
%}

\subsection{The thermal conductivity}
We also want to stress that the thermal conductivity, being essential for
the cooling of young objects such as Cas~A \cite{Blaschke:2011gc}, strongly
depends on in-medium effects like the Landau damping effect in the electron
term and effect of an increase of the $NN$ interaction amplitude with
the density owing to MOPE.
These effects are now consistently included in our calculation scheme.

The thermal conductivity $\kappa$ is given by the sum of partial
contributions, $\kappa = \kappa_b +\kappa_{l}$, where $\kappa_b$
is the heat conductivity of baryons (mainly neutrons) and
$\kappa_{l}$, of leptons (mainly electrons).
The $ep$ crossing term entering $\kappa_l$ proves to be small
\cite{Shternin:2007ee}.
In \cite{Blaschke:2004vq} and \cite{Yakovlev:2003qy} the electron and nucleon
thermal conductivities were  computed according to the analysis of
\cite{Baiko:2001cj}.
More recent studies \cite{Shternin:2007ee} showed that the lepton thermal
conductivity is reduced by an order of magnitude.
Moreover, as we argued in \cite{Blaschke:2004vq},  pion softening effects may
additionally suppress the baryon contribution $\kappa_b$ to the thermal
conductivity.

The impact of a low thermal conductivity on the thermal evolution of neutron
stars accomplished  by introducing  a factor $\zeta_\kappa =0.3$ was first
demonstrated in  Fig.~17 of \cite{Blaschke:2004vq}.
The net effect is a delay of the temperature decline for young ($\lsim 300$~yr)
neutron stars.
This idea of a possible strong suppression of the thermal conductivity
allowed for the explanation of the rapid cooling of Cas~A in
\cite{Blaschke:2011gc}.
In the given work we use the lepton contribution to the thermal conductivity
from \cite{Shternin:2007ee}, cf. Eqs. (40) and (93) of that work.
One may parameterize the result of \cite{Shternin:2007ee} as
 \begin{eqnarray}\label{kappae}
&&\kappa_e = 8.5 \cdot 10^{21} \left(\frac{p_{{\rm F},e}}{{\rm
fm}^{-1}}\right)^{2}f_e\,{\rm ergs}\,\, {\rm s}^{-1} {\rm cm}^{-1}
{\rm K}^{-1}\,, \\ &&f_e \simeq
\frac{2.7}{e^{1.3T/T_{cp}}-1}\,,\nonumber
 \end{eqnarray}
for $T<T_{cp}$ and $f_e =1$ for $T>T_{cp}$.  
For simplicity a  contribution of muons is neglected.

In order to include the effect of the softening of the pion exchange
on $\kappa_b$ we recalculate the $S_{12}$ factors of \cite{Baiko:2001cj}
first with FOPE and then with MOPE and from their ratio we construct an
extrapolation  for $\kappa_b$, which takes into account of the pion softening
effect for $n>n_{cr}^{(1)}$.
Finally we replace
 \be\label{kappan} \kappa_b =\kappa_b^{\rm SY} \left(\omega^*
 (n)/m_{\pi}\right)^3\left(\Gamma (n_0)/\Gamma (n)\right)^4
 n_0/n\,,
 \ee
where $\kappa_b^{\rm SY}$ is  the result of
\cite{Shternin:2007ee}. Note that the main contribution to the
thermal conductivity comes from electrons and we could use
$\kappa_b^{\rm SY}$ for the baryons to get appropriate fit of Cas A. 
We introduce a suppression of $\kappa_b^{\rm SY}$  since it is
in a line with our general argumentation about softening of the
$NN$ interaction owing to the in-medium pion exchange.

\subsection{Blanketing envelope}
In our model the processes occurring in dense neutron star matter
are typically much more efficient than those considered within
minimal cooling modeling. Thus within our scenario  the cooling is
mainly determined by the reactions in the neutron star interior
and much less sensitive to the modeling of the crust. The presence
of a pasta phase \cite{Maruyama:2005vb} at $0.3 \lsim n/n_0 \lsim
0.8$ could partially influence the cooling \cite{Newton:2013zaa}
and the heat conduction due to the possibility of efficient
DU-like neutrino processes. These might be occurring with the
participation of non-uniform structures \cite{Leinson:1993du}, in
spite of the fact that the free proton fraction disappears in
pasta. However, processes in this phase are badly studied. Thereby
this phase is continuing to be ignored in the cooling simulations.
Thus due to ambiguities of its description we ignore the
possibility of the presence of a  pasta layer in our calculations.

The cooling curves essentially depend on the relation between the
internal and surface temperature at small densities $\sim
10^{-3}n_0$ in the blanketing envelope. This relation is not
unique and depends on the mass $\Delta M$ and the structure of the
blanketing envelope. In our works  we use the same band for $T_{s}
- T_{in}$, as it was presented in \cite{Yakovlev:2003ed}. In
\cite{Blaschke:2004vq,Grigorian:2005fn} we demonstrated the
dependence of the cooling curves on  different choices of fits
$T_{s} =f(T_{in})$ within the band and then focused on a model
called ``our fit'', which assumes that for cold more massive
neutron stars the parameter $\eta =\Delta M/M$ is smaller than for
hotter less massive stars. Thus we use a fitting curve $T_{s}
=f(T_{in})$ matching two regimes between $\eta =4\cdot 10^{-16}$
and $\eta = 4\cdot 10^{-8}$. This  ``our fit'' curve is rather
close to a known simplified Tsuruta law. Note that for the mass
$\sim 1.5~M_{\odot}$ ``our fit'' model yields an $\eta$ value
similar to the one used to explain Cas A cooling in
\cite{Page:2010aw}.

\subsection{Equation of state and  massive stars}
As we have mentioned, we adopted the HHJ($\delta =0.2$) EoS for
the description of the nucleon contribution. The energy density of
nucleons is parameterized  as follows
 \be\label{EN}
 E_N=u n_0\left[m_N+ e_{\rm B} u\frac{2+\delta -u}{1+\delta  u} +
 a_{\rm sym}u^{0.6}(1-2x_p)^2\right]\,,
  \ee
where $u=n/n_0$, $e_{\rm B}\simeq -15.8$ MeV is the nuclear
binding energy per nucleon, $a_{\rm sym}\simeq 32$ MeV is the
symmetry energy coefficient and we chose $\delta =0.2$. With these
values of parameters one gets the best fit of APR (A18+$\delta
v$+UIX$^*$)   EoS for symmetric nuclear matter up to $n\sim 4~n_0$.
%{\red
The nucleon effective masses are taken as for the APR EoS.
%}
As we have mentioned, with such $E_N$ one reaches the value of the
maximum mass of the neutron star $M_{\rm max}=1.94~M_{\odot}$, less than
recently measured values of masses of pulsars PSR J1614-2230 and
PSR J0348-0432. The value of the maximum mass can be easily
increased within the HHJ approach provided one diminishes the
parameter $\delta$. However, then one spoils the best fit to the
microscopic APR (A18+$\delta v$+UIX$^*$) EoS. To preserve the best
fit to the APR (A18+$\delta v$+UIX$^*$) EoS for $n\lsim 4 n_0$ we
should perform modifications of EoS only for $n>4n_0$. To do this
we exploit the idea of an excluded volume, related to the quark
substructure of the nucleons and the Pauli exclusion principle on
the quark level. Thus, making use of the replacement
 \be\label{extrapolation}
 u\to \frac{u}{1-\alpha ue^{-(\beta/u)^\sigma}}
  \ee
 in the expression for the energy per particle (in squared
bracket of Eq. (\ref{EN})) we incorporate the excluded volume
effect. We take $\alpha \equiv n_0 v_0 =0.02$, $\beta =6$ and
$\sigma =4$, where $v_0$ has the meaning of an excluded volume.
Thus we derive a new phenomenological HDD EoS. Note that our
parameter choice corresponds to the radius of the  quark core of
the nucleon $r_q \simeq 0.2...0.3$~fm.

In principle pion condensation, if it occurs for $n>n_{cr}^{\pi}$,
softens the EoS. However,  the value of this softening is
essentially model dependent. Therefore, to diminish the dependence
of the EoS on unknown parameters we disregard a possible influence
of pion condensation on the EoS. Also, we suppress possible effect
of hyperonization on EoS.

%%%%%%%%%%%%%%%%%% Figure 3  %%%%%%%%%%%%%%%
\begin{figure}[tbh]
   \includegraphics[width=0.5\textwidth]{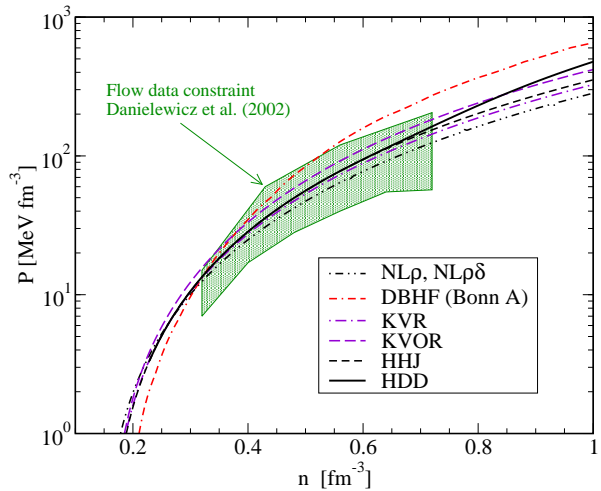}
   \caption{(Color online) Pressure as function of density  consistent with
   a constraint extracted from experimental flow data in isospin symmetric
   nuclear matter \cite{Danielewicz:2002pu} (dark shaded region).
   The HDD EoS (bold  solid line) is introduced in this work, while other
   EoSs
   shown for comparison are used according to Ref.~\cite{Klahn:2006ir} and the
   notation introduced there.
    }
   \label{dan_expol_HDD}
\end{figure}
%%%%%%%%%%%%%%%%%%%%%%%%%%%%%%%%%%

%%%%%%%%%%%%%%%%%%% Figure 4 %%%%%%%%%%%%%%
\begin{figure}[tbh]
   \includegraphics[width=0.5\textwidth]{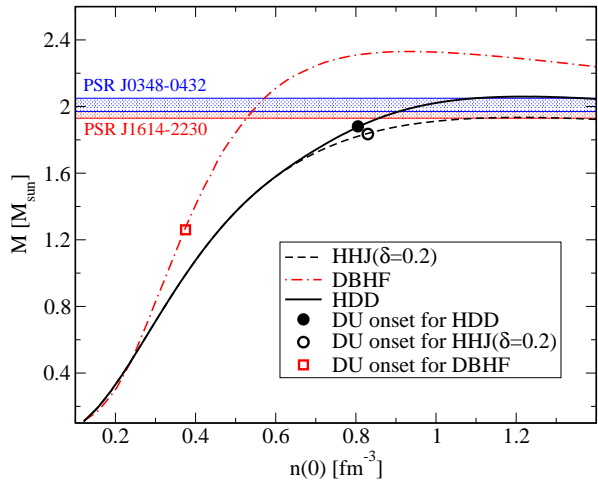}
   \caption{(Color online) Neutron star mass vs. central density for
   HHJ($\delta =0.2$) (dashed line), DBHF (dash-dotted line)
   and HDD (bold solid line) EoSs.
   Symbols on the lines indicate the thresholds of the DU reaction.
   }
   \label{HHJ-NLW-DU0nsets}
\end{figure}
%%%%%%%%%%%%%%%%%%%%%%%%%%%%%%%%%%

In Fig.~\ref{dan_expol_HDD} we compare different EoS models for
isospin-symmetric nuclear matter with a constraint region
extracted from an analysis of experimental flow data
\cite{Danielewicz:2002pu}. It is seen that both, the HHJ($\delta
=0.2$) and the HDD EoSs satisfy the experimental constraint
(shaded region). The HDD EoS stiffens only for $n>4n_0$. Causality
is not violated  up to the limiting central density for stable
neutron stars.

In Fig.~\ref{HHJ-NLW-DU0nsets} we show neutron star masses vs.
central density $n(0)$ for HHJ($\delta =0.2$), DBHF and HDD EoSs.
With our parameter choice the HDD EoS produces a maximum mass
$M_{\rm max}= 2.06~M_{\odot}$. The DU threshold is changed only
slightly compared to that for the HHJ($\delta =0.2$) EoS while
DBHF has a very early DU onset.

Note that our HDD EoS might be rather convenient for the
description of possible phase transitions, like the hadron --
quark phase transition, in massive neutron stars. The latter
transition may occur as a first or second order phase transition,
as a crossover, or as the melting of hadron matter, when the quark
cores of hadrons become essentially overlapping. At $n> 4 n_0$ the
hadron pressure additionally increases compared to that for HHJ
EoS, which favors the transition to quark matter at a smaller
pressure for fixed baryon chemical potential.

\section{Neutron Star in Cas A}
%{\it The Neutron Star in Cas A.--}
The ingredients of the nuclear medium cooling scenario discussed above lead
to neutron star cooling curves in Fig.~17 of Ref.~\cite{Blaschke:2004vq},
where model I for the proton gap has been adopted and the role of the
thermal conductivity on the hot early stages of hadronic neutron
star cooling was elucidated (see curves for $\kappa =0.3$ in
\cite{Blaschke:2004vq}).
In Fig.~1 of \cite{Blaschke:2011gc} we redrew those cooling curves permitting
readjustment of the thermal conductivity parameter.
This allowed us to describe all cooling data including those for Cas A, known
at that time.

%%%%%%%%%%%%%%%% Figure 5 %%%%%%%%%%%%%%%%%
\begin{figure}[tbh]
   \includegraphics[width=0.50\textwidth]{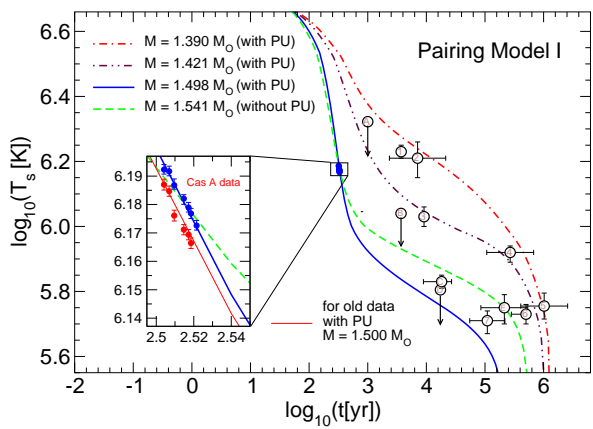}
   \caption{(Color online) Cooling of neutron stars
     within
     nuclear medium cooling scenario, with
     PU and without PU for model I for pairing,
%     for three different stellar masses, $M$,
      cf. also Fig.~17 of \cite{Blaschke:2004vq} and Fig.~1 of
     \cite{Blaschke:2011gc}.
     Data  from Refs.~\cite{Page:2004fy,Yakovlev:2010ed}.
     New data for Cas A (above the old ones) from \cite{Elshamouty:2013nfa}.
}
   \label{CasA_APR(HHJ)_1}
\end{figure}
%%%%%%%%%%%%%%%%%%%%%%%%%%%%%%%%%%
In Fig.~\ref{CasA_APR(HHJ)_1} we show results of our new calculation for the
model I for pairing, with pion condensation (with PU), see curves 1a+2 and 3
in Fig.~\ref{omegatil}, and without pion condensation (without PU), see
curves 1a+1b  in Fig.~\ref{omegatil}.
As we mentioned, we use {\it all} the same parameters, as in our previous
calculation  \cite{Blaschke:2011gc} but instead of an {\it ad hoc} suppression
of $\kappa$ we now exploit calculated values of $\kappa$, cf. Eqs.
(\ref{kappae}), (\ref{kappan}) above.
The change of $\kappa$ affects the cooling of young objects ($t\lsim 300$ yr)
only.

With PU we perfectly explain the new Cas A data for $M=1.498~M_{\odot}$
and the old Cas A data for $M=1.500 M_{\odot}$ (in \cite{Blaschke:2011gc} we
had $M=1.463~M_\odot$ and $\zeta_\kappa = 0.265$).
As is seen from Fig.~\ref{CasA_APR(HHJ)_1}, with the neutron star mass
$M =1.390~M_\odot$  the upper data points are covered.
The lower data points are covered for the mass of Cas A.
Thus the whole set of available cooling data is covered with masses in the
range $1.390<M/M_\odot<1.5~$.

Assuming the absence of  pion condensation in the core of a neutron
star (model ``without PU''), the new Cas~A cooling data are described for
$M=1.541~M_{\odot}$ whereas for $M=1.500~M_{\odot}$ the old Cas A data are
reproduced (in \cite{Blaschke:2011gc} we had for this  case $M=1.532~M_\odot$
and $\zeta_\kappa = 0.175$).

%%%%%%%%%%%%%%%%% Figure 6 %%%%%%%%%%%%%%
\begin{figure}[!h]
  \includegraphics[width=0.50\textwidth]{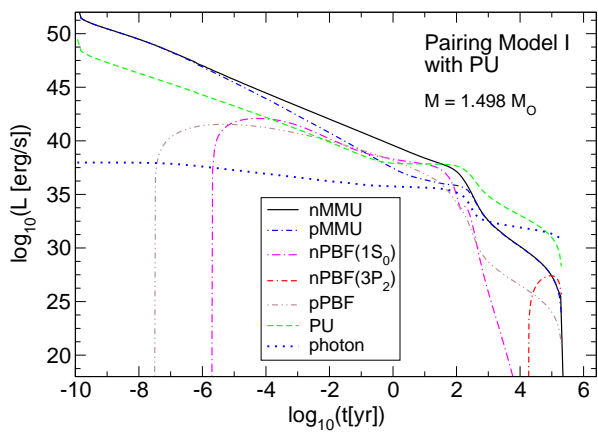}
  \caption{(Color online) Individual contributions of the
    cooling processes, nMMU and pMMU, $1S_0$ pPBF and nPBF, $3P_2$
    nPBF, PU, and surface photon emission, to the total stellar
    luminosity for the neutron star of mass $1.498~M_{\odot}$  shown in
    Fig.~\ref{CasA_APR(HHJ)_1}.
    Modeling with PU exploits curves 1a+2 for the pion excitations, and 3
    determining the pion condensate amplitude in Fig.~\ref{omegatil}, pairing
    gaps  from model I. }
   \label{CasA_Lt_piY}
\end{figure}
%%%%%%%%%%%%%%%%%%%%%%%%%%%%%%%

%%%%%%%%%%%%%%%%% Figure 7 %%%%%%%%%%%%%%
\begin{figure}[!h]
  \includegraphics[width=0.5\textwidth]{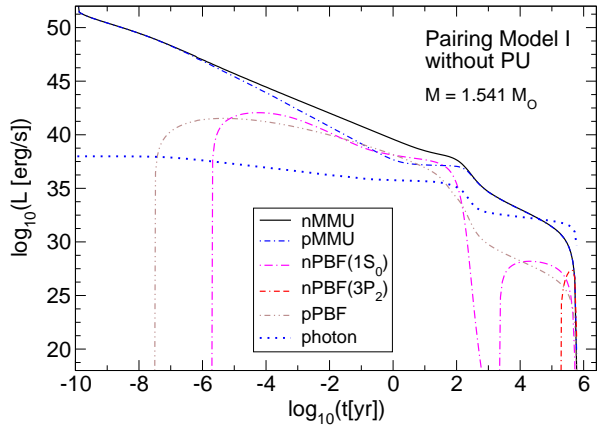}
  \caption{(Color online) The same as in Fig.~\ref{CasA_Lt_piY}
  but for the model ``without PU''.
   }
   \label{CasA_Lt_Y}
\end{figure}
%%%%%%%%%%%%%%%%%%%%%%%%%%%%%%%

In Fig.~\ref{CasA_Lt_piY} we show the individual contributions of
the cooling processes in our scenario to the total neutron star
luminosity for
%our best- fit
the neutron star with the mass $M = 1.498~M_{\odot}$, which best
reproduces the new cooling data of Cas~A in
Fig.~\ref{CasA_APR(HHJ)_1}, for model I for pairing   and PU
model. We see that the nMMU and PU are the most efficient
processes in this scenario, while $1S_0$ PBF processes are less
important. Nevertheless they contribute  at $t\lsim 500$ yr.
Although the MnB and MpB luminosities dominate over those of PBF
in a broad time interval, they are not shown since they have
rather similar shapes, as the nMMU and pMMU curves, but  have
smaller amplitudes. Note that PU processes control the neutron
star cooling at later times ($300\lsim t/{\rm yr} \lsim 10^5$).
For $t> 10^5$ yr the photon emission from the star surface is
dominating. The $3P_2$ nPBF process is unimportant with values of
the gap that we use.

 In Fig.~\ref{CasA_Lt_Y} we show the individual contributions of
the cooling processes in our scenario to the total neutron star
luminosity for
%our best- fit
the neutron star with mass $M = 1.541~M_{\odot}$, which best
reproduces the new cooling data of Cas~A in
Fig.~\ref{CasA_APR(HHJ)_1} for model I for pairing  and without
PU. We see that the nMMU  are the most efficient processes in this
scenario during the whole time evolution up to $t\lsim 10^5$ yr. A
little jump in the luminosity on the $1S_0$ nPBF at $10^3\lsim
t/{\rm yr} \lsim 10^5$ is due to a still surviving inhomogeneity
of the temperature profile at low densities, in the crust. This
tiny effect does not influence the neutron star cooling evolution.
It completely disappeared in the above considered PU model, where
the mentioned temperature inhomogeneity is smoothed faster owing
to more efficient heat transport.

\section{Cooling of neutron stars with different masses}
\begin{figure}[tbh]
   \includegraphics[width=0.50\textwidth]{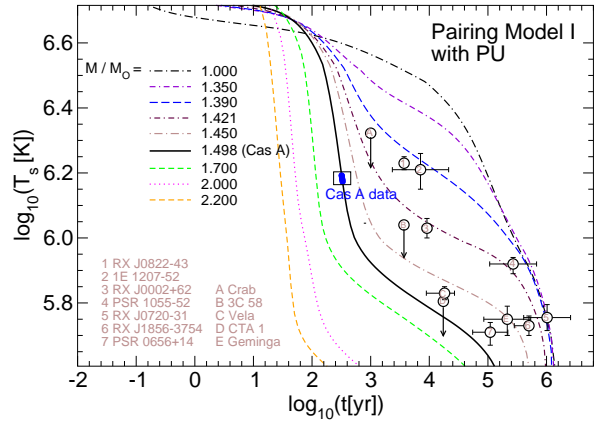}
   \caption{(Color online)
     Cooling of neutron stars in the nuclear medium cooling scenario, with
     different masses within model I for pairing and with PU,
%     for three different stellar masses, $M$,
     cf. also Fig.~17 of \cite{Blaschke:2004vq} and Fig.~1 of
     \cite{Blaschke:2011gc}.
     Data  from Refs.~\cite{Page:2004fy,Yakovlev:2010ed}.
     New data for Cas A from \cite{Elshamouty:2013nfa}.
}
   \label{CasA_APR(HHJ)_3}
\end{figure}

\begin{figure}[tbh]
   \includegraphics[width=0.50\textwidth]{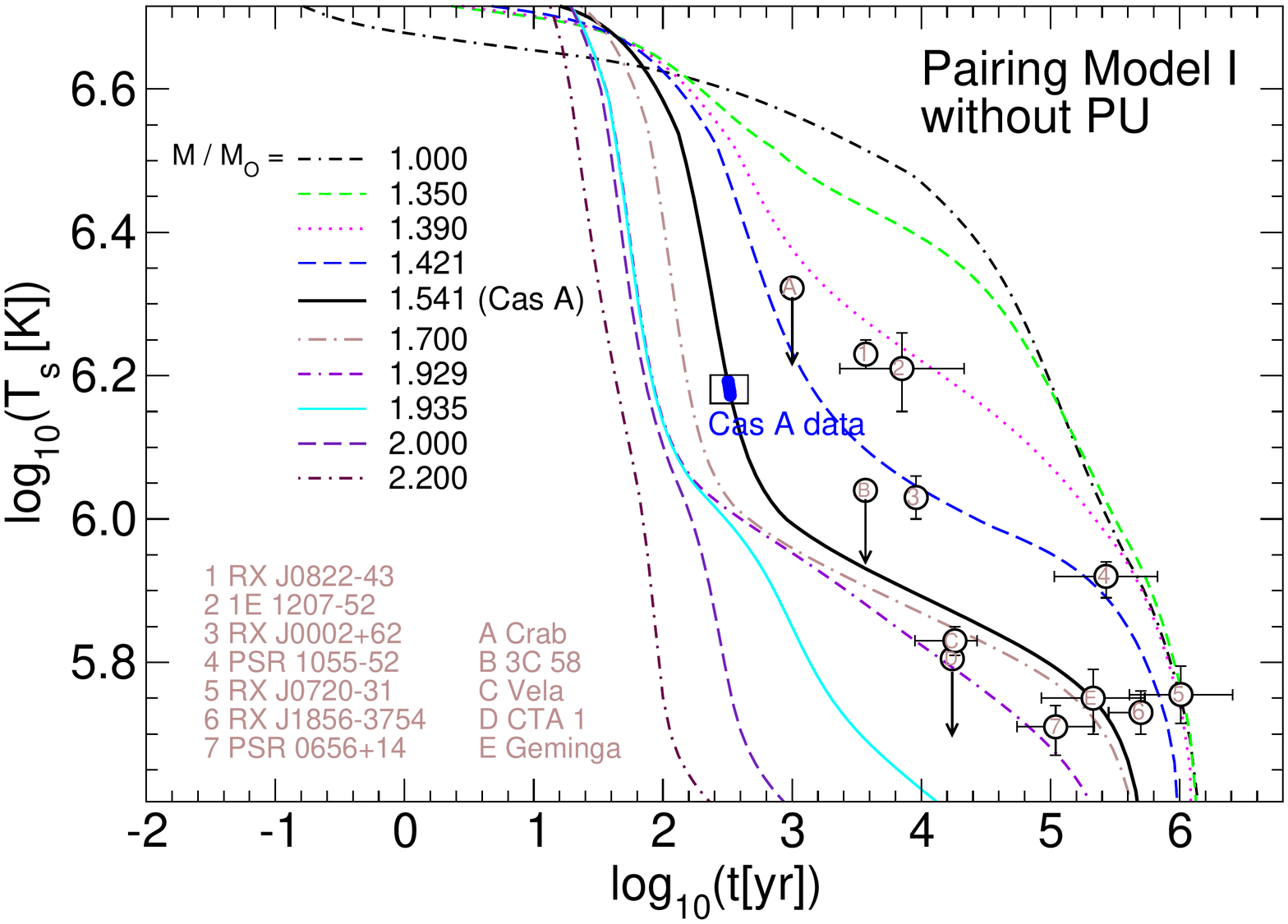}
   \caption{(Color online)
    The same as in Fig. \ref{CasA_APR(HHJ)_3} but for the model without PU.
}
   \label{CasA_APR(HHJ)_2}
\end{figure}

In Figs.~\ref{CasA_APR(HHJ)_3} and \ref{CasA_APR(HHJ)_2} we
demonstrate the general picture of the cooling of neutron stars
with different masses for the model I for pairing.
Fig.~\ref{CasA_APR(HHJ)_3} shows cooling in the model, where  PU
processes are included, whereas Fig.~\ref{CasA_APR(HHJ)_2},
without PU. Although the overall picture is similar, in the model
with PU presently available data are explained within an
essentially narrower interval of neutron star masses ($1.39
<M/M_{\odot}<1.5$) than in the model without PU ($1.39<M/
M_{\odot}<1.9$). The cooling of stars with  $M> M_{\rm DU} =
1.881~M_{\odot}$  is controlled by the efficient DU process. The
heaviest stars, which we now are able to describe with the HDD
EoS, are cooled so fast that they are not seen in soft $X$-rays at
least at present.
%
%%%%%%%%%%%%% Figure 8 %%%%%%%%%%%%%%%%%%%%
%%%%%%%%%%%%%%%%%%%%%%%%%%%%%%%%%%

%%%%%%%%%%%%%%% Figure 9 %%%%%%%%%%%%%%%%%%
%%%%%%%%%%%%%%%%%%%%%%%%%%%%%%%%%%

%%%%%%%%%%%%%%% Figure 10 %%%%%%%%%%%%%%%%%%
\begin{figure}[tbh]
   \includegraphics[width=0.50\textwidth]{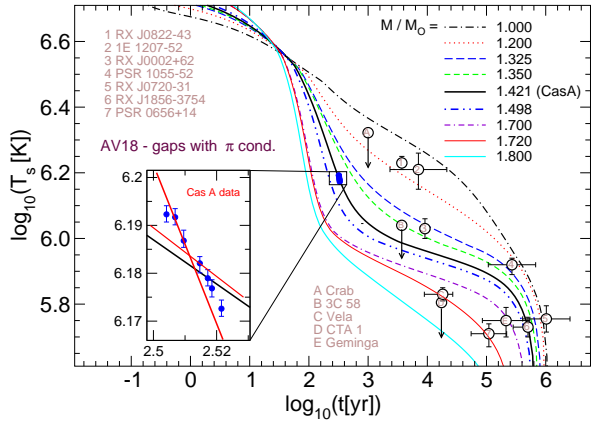}
   \caption{(Color online)
    The same as in Fig. \ref{CasA_APR(HHJ)_3} but for the model II for pairing.
    Two extra lines in the inset frame show $2\%$ and $5.5\%$ declines.
}
   \label{CasA_APR(HHJ)_5}
\end{figure}
%%%%%%%%%%%%%%%%%%%%%%%%%%%%%%%%%%
Fig.~\ref{CasA_APR(HHJ)_5}  shows the cooling of neutron stars
with different masses for the model II for pairing and for the
model with PU. The  $4\%$ decline is not fitted without additional
strong suppression of $\kappa$ (in agreement with statement of
\cite{Blaschke:2011gc} where we needed to suppress the thermal
conductivity by factor $\zeta_{\kappa} \leq 0.015$ in order to
describe a $4\%$ decline with model II). However, the
cooling curve corresponding to the star mass $M=1.421~M_{\odot}$
matches the $2\%$ decline (upper experimental border). Thus the
scenario using model II still cannot be excluded
at present. Subsequent measurements may allow to reduce the
uncertainty in the decline, which will allow to distinguish
between our scenarios using models I and II for pairing.

\section{Summary and Conclusion}
We have shown in this paper that the nuclear medium cooling scenario allows
one to nicely explain the observed rapid cooling of the neutron star in Cas~A,
as well as all other existing neutron star cooling data.
As demonstrated already in \cite{Blaschke:2004vq} and then in
\cite{Blaschke:2011gc}, in our scenario the rapid cooling of very young
objects like Cas~A is mainly due to the efficient MMU processes, a very low
(almost zero) value of the $3P_2$ neutron gap
%%, a large proton gap
and a small thermal conductivity of neutron star matter.
In the present work we do not use any artificial suppression parameter to
demonstrate the effect of a small thermal conductivity caused by in-medium
effects on Cas A cooling, but we {\it use the same values for the lepton
thermal conductivity} as in Ref.~\cite{Shternin:2007ee}.
The required smallness of thermal conductivity is provided by
taking into account the collective  effect of Landau damping.

We stress that, contrary to the minimal cooling models, within the
nuclear medium cooling scenario we are trying to consistently
include the most important collective effects in all relevant
processes -- the pion softening effect in the $NN$ interaction
amplitude \cite{Migdal:1990vm}, collective effects in the
pair-breaking-formation processes \cite{Kolomeitsev:2008mc},
collective effects in the pairing gaps, the screening effect in
the lepton contribution to the thermal conductivity
\cite{Heiselberg:1993cr} and a decrease of the nucleon
contribution owing to the mentioned pion softening, etc. And we
did not introduce any significant changes in  our scenario
developed in 2004 in \cite{Blaschke:2004vq}, except for including
the suppression of the thermal conductivity, now performed as in
\cite{Shternin:2007ee}. Thus in difference with other scenarios,
explanation of  the Cas A data  straightly follows the predictions
of our previous work \cite{Blaschke:2004vq}.

The pion softening effect manifesting itself in an increase of the
$NN$ interaction amplitude with growing density
\cite{Migdal:1978az,Migdal:1990vm} causes a decrease of  the
nucleon contribution to the thermal conductivity and at the same
time leads to a strong enhancement of the emissivities of
two-nucleon neutrino processes
\cite{Voskresensky:1986af,Migdal:1990vm}. With due  account of
exact vector current conservation the pair-breaking-formation
processes on the vector current prove to be dramatically
suppressed \cite{Leinson:2006gf} and operate instead on the axial
current \cite{Kolomeitsev:2008mc}, for which the suppression
effect is less pronounced. Following calculations of
\cite{Schwenk:2003bc} which take into account polarization
effects, the $3P_2$ gap is dramatically suppressed. The screening
effect taken into account in the calculation of the lepton thermal
conductivity \cite{Shternin:2007ee} leads to its strong
suppression compared to the earlier result of
Ref.~\cite{Baiko:2001cj}. We included only the most efficient
in-medium processes. However, the results are sensitive to details
of the description of strong interactions in dense matter. Thus
demonstrating nice agreement with the data we only argue in favor
of a general picture but not  guarantee  quantitative values of
all involved quantities. There are still many other in-medium
reaction channels which we did not include in the code.
For example, in superfluid matter spin excitonic and diffusive modes
\cite{Kolomeitsev:2011wz} and massive photon decay
\cite{Voskresensky:1998nk} may contribute, while in pasta phases
DU-like processes on structures \cite{Leinson:1993du} may operate,
%{\red
the nucleon Fermi sea may be rearranged in the vicinity of
the pion condensation point \cite{KhodelVoskr}.
We did not consider possibilities of other phase transitions except
pion condensation.
%}
Unfortunately, quantitative estimates of these processes depend
on the values of some not well known parameters.
Therefore, we postpone their inclusion in the cooling
code in order not to multiply uncertainties.

Thus our explanation of the Cas~A cooling constitutes an
alternative to that of
\cite{Page:2010aw,Shternin:2010qi,Elshamouty:2013nfa}, which is
based on a strong nPBF process due to $3P_2$ superfluidity in
neutron star interiors and suppressed emissivities of pPBF and MU
two-nucleon processes operating on the charged current by a
suggestion to use a large proton gap. The conclusion that from Cas
A observations one is able to recover the value of the $3P_2$
$nn$-pairing gap seems to us misleading, due to the existence of
other (not  exotic!) possibilities to explain Cas~A, as well as
all other available cooling data. We support, however, the
conclusion of these authors
%that a large value of the proton gap
%is preferable, albeit not necessarily in the entire neutron star core.
about the sensitivity of the result to the chosen value of the proton gap.
In our scenario this occurs due to the sensitivity of the  MMU emissivity to
the value and the density dependence  of the proton gap spreading up to
$n\lsim 3\dots 4 n_0$ in the neutron star core, where the MMU process is most
efficient.
We got the best ($4\%$ decline) fit of Cas A data with a larger proton gap
(model I).
Nice overall agreement with available cooling data is achieved with a tiny
$3P_2$ $nn$ pairing gap and it would be destroyed, if we used values of the
$3P_2$ pairing gap similar to those used in
\cite{Page:2010aw,Shternin:2010qi,Elshamouty:2013nfa}.

Alternative explanations include the suggestion \cite{Negreiros:2011ak} that
Cas~A is a rapidly rotating star and during its spin-down the efficient DU
process is switched on when the redistribution of matter leads to an increase
of the central density beyond the DU threshold.
Ref.~\cite{Sedrakian:2013pva} suggests that Cas~A is a hybrid star.
However, note that except our scenario other works aiming at a description of
Cas~A do not demonstrate their capability to obtain in the framework of the
same assumptions an overall agreement with other available neutron star
cooling data.

Further tests may be considered, such as a comparison of log N-log S
distributions from population synthesis with the observed one for isolated
neutron stars \cite{Popov:2004ey}.
Also a continuation of the measurements of Cas~A and new measurements of
neutron star temperatures are welcome to discriminate between alternative
cooling scenarios.

In this paper we also incorporated an excluded volume effect in
the HHJ($\delta=0.2$) EoS thus extending our previous works in
order to describe the cooling of stars with masses $\gsim
2~M_{\odot}$. We demonstrated that a difference of the here
constructed HDD EoS with the HHJ($\delta=0.2$) EoS appears only
for densities $>4 n_0$. Thereby only the cooling curves starting
from  $M\sim 1.6~M_{\odot}$ are affected by this change of the
EoS.
The HDD EoS might be helpful to study  hybrid stars, owing to
its additional stiffening  at densities exceeding $4 n_0$, when a
phase transition to quark matter is expected \cite{Klahn:2013kga}.
We hope to return to this analysis in a forthcoming publication.

\subsection*{Acknowledgments}
We thank E.~E.~Kolomeitsev and F.~Weber for discussions.
This work was supported by Narodowe Centrum Nauki under contract
No. DEC-2011/02/A/ST2/00306 and by
CompStar, a research networking programme of the European Science Foundation.
D.B. has been supported by RFBR under grant No. 11-02-01538-a and
H.G. acknowledges support by the Volkswagen Foundation under grant No. 85 182.

\end{document}